# A DESIGN OF A FAST PARALLEL-PIPELINED IMPLEMENTATION OF AES: ADVANCED ENCRYPTION STANDARD

Ghada F.Elkabbany, Heba K.Aslan and Mohamed N.Rasslan

Informatics Department, Electronics Research Institute, Cairo, Egypt

## ABSTRACT

*The Advanced Encryption Standard (AES) algorithm is a symmetric block cipher which operates on a sequence of blocks each consists of 128, 192 or 256 bits. Moreover, the cipher key for the AES algorithm is a sequence of 128, 192 or 256 bits. AES algorithm has many sources of parallelism. In this paper, a design of parallel AES on the multiprocessor platform is presented. While most of the previous designs either use pipelined parallelization or take advantage of the Mix_Column parallelization, our design is based on combining pipelining of rounds and parallelization of Mix_Column and Add_Round_Key transformations. This model is divided into two levels: the first is pipelining different rounds, while the second is through parallelization of both the Add_Round_Key and the Mix_Column transformations. Previous work proposed for pipelining AES algorithm was based on using nine stages, while, we propose the use of eleven stages in order to exploit the sources of parallelism in both initial and final round. This enhances the system performance compared to previous designs. Using two-levels of parallelization benefits from the highly independency of Add_Round_Key and Mix_Column/ Inv_Mix_Colum transformations. The analysis shows that the parallel implementation of the AES achieves a better performance. The analysis shows that using pipeline increases significantly the degree of improvement for both encryption and decryption by approximately 95%. Moreover, parallelizing Add_Round_Key and Mix_Column/ Inv_Mix_Column transformations increases the degree of improvement by approximately 98%. This leads to the conclusion that the proposed design is scalable and is suitable for real-time applications.*

## KEYWORDS

*Advanced Encryption Standard AES, Parallel processing, Pipelining*

## 1. INTRODUCTION

On June 2, 1997, the American National Institute for Standardization and Technology (NIST) proposed a competition to propose a new encryption algorithm to replace the aging and increasingly vulnerable Data Encryption Standard (DES). The new Advanced Encryption Standard (AES) chosen from the competitors was Rijndael [1 and 2]. Since becoming the AES, Rijndael has been the focus of countless analyses and has been implemented both in hardware and software for many different platforms. To accelerate the AES computation time, parallel computing is incorporated [3-19].

In this paper, a design of parallel AES on the multiprocessor platform is presented. While most of the previous designs either use pipelined parallelization or take advantage of the Mix_Column parallelization, our design is based on combining pipelining of rounds and parallelization of Mix_Column and Add_Round_Key transformations. This model is divided into two levels: the first one is pipelining different rounds, while the second one is through parallelization of both the Add_Round_Key and the Mix_Column transformations. Previous work proposed for pipelining AES algorithm was based on using nine stages, while, we propose the use of eleven stages in order to exploit the sources of parallelism in both initial and final round. The paper is organized as follows: in Section 2, a description of AES algorithm and a survey of different designs for its implementation in parallel are detailed. Then, the proposed design is illustrated in Section 3. In Section 4, a performance evaluation of the proposed design is given. Finally, the paper concludes in Section 5.

# 2. RELATED WORK

## 2.1. Advanced Encryption Standard (AES)

The Advanced Encryption Standard (AES) algorithm is a symmetric block cipher which can convert data to an unintelligible form (encryption) and convert the data back into its original form (decryption). Both encryption and decryption consist of sequences of blocks each consists of 128-bits. Moreover, the cipher key for the AES algorithm is a sequence of 128, 192 or 256 bits. Internally, the AES algorithm's operations are performed on a two-dimensional (2-D) array of bytes called the *State* array. The *State* array consists of four rows of bytes, each containing "$N_b$" bytes, where "$N_b$" is the block length divided by 32 (the word size).

### Description of the AES Algorithm

The AES algorithm consists of three distinct phases as shown in Figure 1 [3]:

- In the first phase, an initial addition (XORing) is performed between the input data (plaintext) and the given key (cipher key).
- Then, in the second phase, a number of standard rounds (*Nr*-1) are performed, which represents the kernel of the algorithm and consumes most of the execution time. The number of these standard rounds depends on the key size; nine for 128-bits, eleven for 192-bits, or thirteen for 256-bits. Each standard round includes four fundamental algebraic function transformations on arrays of bytes namely:

    (1) Byte substitution using a substitution table (Sbox)
    (2) Shifting rows of the *State* array by different offsets (ShiftRow)
    (3) Mixing the data within each column of the *State* array (Mix_Column), and
    (4) Adding a round key to the *State* array (Key-Addition).

- Finally, the third phase of the AES algorithm represents the final round of the algorithm, which is similar to the standard round, except that it does not have Mix_Column operation. For detailed information of the abovementioned transformations, the reader could refer to [1].

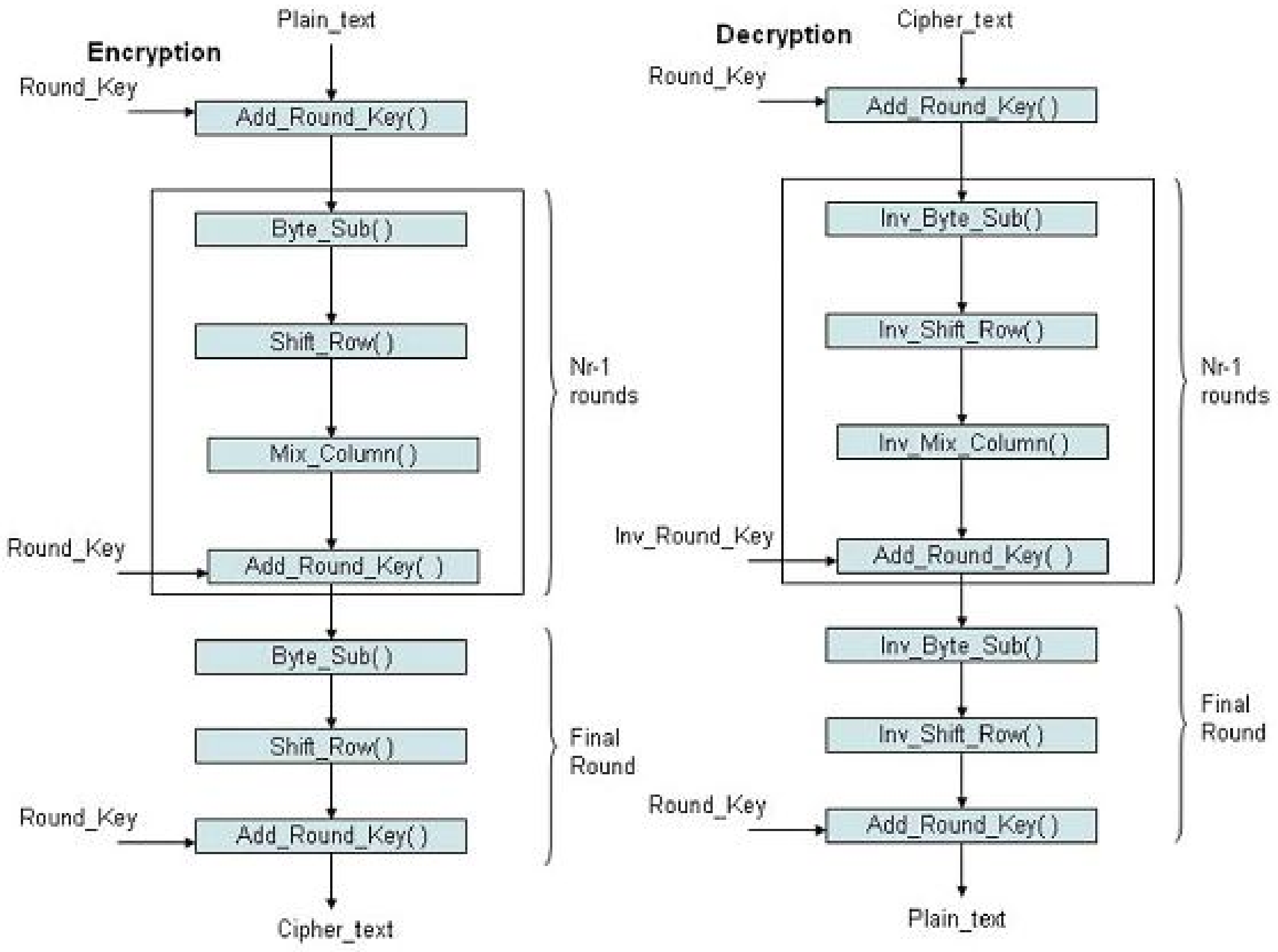

Figure 1. The AES algorithm (Nr: 10, 12, or 14 depending on key length) [4]

## 2.2. The Parallel Advanced Encryption Standard (AES)

Advanced Encryption Standard (AES) can be deployed in fully hardware [3-11], hybrid software-hardware [12-16], and fully software implementations [17-19]. This fact allows parallelization of AES in different ways. In literature, parallelizing Rijndael has been visited many times for hardware implementation. In [4], Yoo et al. presented a hardware-efficient design that increases AES throughput by making use of a high-speed parallel pipelined architecture. Yoo et al. used an efficient inter-round and intra-round pipeline design in order to achieve a high throughput in encryption. In each round, there are three pipeline stages, the first stage follows the byte-sub operation, the second one is located after the shift-row operation, and the last stage is before data output. Moreover, this design has one pipeline stage in key generation blocks. On the other hand, Hodjad el at. [5] introduce a design that has four or seven pipeline stages, one after a byte-sub operation and three or six in a byte-sub operation. In [7], Ananth et al. present a fully pipelined AES encryption/decryption system that is fully unrolled in order to implement a very deep level of pipelining (i.e. all ten cipher rounds were unrolled.) For more designs for hardware implementation of AES, the reader could refer to [8-11].

On the other hand, AES in software-hardware co-design is performed by using extended special instructions and the other transforms are performed by general instructions [12-16]. S. Mahmoud [12] presented a parallel implementation for AES algorithm by using the MPI (Message Passing Interface) based cluster approach. MPI is one of the most established methods used in parallel programming mainly. This is due to the fact that the relative simplicity of deploying the method by writing a set of library functions or an API (Application Program Interface) callable from C, C++ or Fortran Programs. In [13], So-In shows that the 16-bytes AES block can be individually encrypted. As an essential technique of AES parallelism is to execute parallel AES by applying each thread or each node into each AES block to establish a complete encrypted parallel block. This technique excludes the key expansion step required before entering the parallel state. So-In applies AES encryption in ECB mode for the sake of performance evaluation. Similarly, CTR mode can be encrypted without the dependency of the previous blocks, but not other modes. Other designs that use instruction set to increase the efficiency of 32-bit processors for AES encryption algorithm could be find in [14-16].

In [17], Brisk et al. introduce an example of fully software implementation of AES. In their work, they derived the asymptotic sequential runtime for the algorithm and describe two parallel implementations. The first one is optimal in terms of time consuming and the other one is optimal in terms of cost. In the cost-optimal implementation, they sacrifice acceleration in order to reduce the number of processors required for encryption. Other examples of fully software implementations are presented in [18 and 19].

In this paper, a design of parallel AES on the multiprocessor platform is presented. While most of the previous designs either use pipelined parallelization or take advantage of the Mix_Column parallelization, in our work, we design a parallel model for the AES algorithm. This model is divided into two levels. The first one is pipelining different rounds, while the second one is through parallelization both Add_Round_Key and Mix_Column transformations. In the next section, the proposed parallel AES design is presented.

## 3. THE PROPOSED PARALLEL ADVANCED ENCRYPTION STANDARD (AES) DESIGN

Advanced Encryption Standard (AES) algorithm has many sources of parallelization as mentioned in Section 2. In this work, we design a parallel model for the AES algorithm, this model is divided into two levels. The first one is pipelining different rounds, while the second one is through parallelization both the Add_Round_Key and the Mix-Column. In this section, the parallel design of the AES algorithm is explained, while in the next section its analysis is detailed.

### 3.1. The Parallel Encryption Model

Based on the AES description in Section 2.1, AES algorithm is divided into three distinct phases. *The first phase* contains the initial round. *The Second phase* contains "*Nr*-1" standard rounds, in which each round includes four transformations namely: Byte_Sub, Shift_Row, Mix_Column, and Add_Round_Key. Finally, *the third phase* contains the final round. That is similar to any standard round, except that it does not have Mix_Column transformation. Both Byte_Sub, and Shift_Row transformations are executed sequentially because they operate on single bytes, independently of their position in the State matrix. On the other hand, Mix_Column and Add_Round_Key operations can be executed in parallel. While the Add_Round_Key operation is used to perform an arithmetic XOR operation, the Mix_Column transformation, which represents the kernel of the AES algorithm and consumes most of the execution time, is used to perform 64 XOR operations and 32 shift operations.

In this work, we design a parallel model for the AES algorithm, this model is divided into two levels. The first one is pipelining different rounds (from round zero to round 10), while the second one is through parallelization of both Add_Round_Key and Mix-Column transformations.

### 3.1.1 Pipelined Encryption Rounds

As shown in Figure 1, round number zero (initial round) through round number "$N_r$, $N_r$=10" represent the individual rounds in the AES-128 encryption. The pipelining between these rounds will achieve a high performance implementation. The data generated in each individual round is used as the input to the next round. This is one of the easiest methods where high performance can be achieved in a very minimal amount of time, thus, reducing the overall design implementation cycle.

We assume that our system contains eleven stages {$S_0$,$S_1$,$S_2$ ,….., $S_9$, and, $S_{10}$}, and the total number of processors equals "*M*". Moreover, each processor has its local memory, and the processor and its memory are called processing element. The *"M"* processing elements connected to each other via multiport Shared Memory (SM). The content of a multi-port memory can be accessed through different ports simultaneously. In our work, each stage can be performed by $M_r$ processing elements PEs, where $M_r = M/11$. Each group of "$M_r$" PEs has a direct independent access to a certain memory module, and each PE has a dedicated path to each module in order to achieve a better performance. On the other hand, different stages are connected through pipelined stream. That is to say, the pipelined stream contains eleven functions each function is executed by a single stage. There is a pipeline stage between each round and the parallelization inside each round which will be described in Section 3.1.2. Our pipeline design is different from [4] by adding two stages $S_0$ and $S_{10}$ to the pipeline stream. Each of these stages are used to execute the Add_Round_Key transformation (this transformation consists of sixteen XOR operations), i.e. the design is fully pipelined. This technique of pipelining will increase the concurrency and reduce the total execution time.

### 3.1.2 Parallelization inside Individual Round

Each individual round consists of four transformations. As mentioned earlier, Mix_Column and Add_Round_key transformations can be executed in parallel. Add_Round_key consists of 16 independent XOR operations, therefore, it could be executed in parallel. In addition, Mix_Column transformation consists of 64 XOR operations and 32 shift operations. Mix_Column represents the kernel of the AES algorithm and consumes most of the execution time. This necessitates its implementation in parallel to reduce its execution time. In this section, the mathematical derivation of Mix_Column is discussed in details. In our design, "E" represents the matrix used for encryption, while "D" represents the matrix used for decryption. On the other hand, we assume that "$B_i$" and "$C_i$" are the input and output of the Mix_Column operation in case of encryption, and are inversed at the decryption process. In order to encrypt *"L"* number of data blocks ( $1 \le i \le L$ ), E, $B_i$, and $C_i$, for each block can be represented as follows:

$$B_i = \begin{bmatrix} b_{i,1} & b_{i,2} & b_{i,3} & b_{i,4} \\ b_{i,5} & b_{i,6} & b_{i,7} & b_{i,8} \\ b_{i,9} & b_{i,10} & b_{i,11} & b_{i,12} \\ b_{i,13} & b_{i,14} & b_{i,15} & b_{i,16} \end{bmatrix} \quad \text{and} \quad E = \begin{bmatrix} 02 & 03 & 01 & 01 \\ 01 & 02 & 03 & 01 \\ 01 & 01 & 02 & 03 \\ 03 & 01 & 01 & 02 \end{bmatrix}$$

$$C_i = E * B_i \tag{1}$$

$$\begin{bmatrix} c_{i,1} & c_{i,2} & c_{i,3} & c_{i,4} \\ c_{i,5} & c_{i,6} & c_{i,7} & c_{i,8} \\ c_{i,9} & c_{i,10} & c_{i,11} & c_{i,12} \\ c_{i,13} & c_{i,14} & c_{i,15} & c_{i,16} \end{bmatrix} = \begin{bmatrix} 02 & 03 & 01 & 01 \\ 01 & 02 & 03 & 01 \\ 01 & 01 & 02 & 03 \\ 03 & 01 & 01 & 02 \end{bmatrix} * \begin{bmatrix} b_{i,1} & b_{i,2} & b_{i,3} & b_{i,4} \\ b_{i,5} & b_{i,6} & b_{i,7} & b_{i,8} \\ b_{i,9} & b_{i,10} & b_{i,11} & b_{i,12} \\ b_{i,13} & b_{i,14} & b_{i,15} & b_{i,16} \end{bmatrix} \tag{2}$$

Mix_Column transformation is then represented by the following set of equations and is illustrated in Figure 2:

$$c_{i,1} = (2 \bullet b_{i,1}) \oplus (3 \bullet b_{i,5}) \oplus b_{i,9} \oplus b_{i,1} \tag{3}$$

$$c_{i,5} = b_{i,1} \oplus (2 \bullet b_{i,5}) \oplus (3 \bullet b_{i,9}) \oplus b_{i,13} \tag{4}$$

$$c_{i,9} = b_{i,1} \oplus b_{i,5} \oplus (2 \bullet b_{i,9}) \oplus (3 \bullet b_{i,13}) \tag{5}$$

$$c_{i,13} = (3 \bullet b_{i,1}) \oplus b_{i,5} \oplus b_{i,9} \oplus (2 \bullet b_{i,13}) \tag{6}$$

This is repeated for the other three columns of the matrix. The above description shows that the elements of Mix_Column matrix can be computed independently. The Mix_Column transformation can be executed by more than one processor. The maximum number of processors is thirty-two processors in each stage. Figure 3 represents the proposed parallel design for the AES encryption operation, while Figure 4 describes the details of computing each matrix element in parallel. As shown in this figure, two processors can cooperate to compute one or more element $c_{i,j}$.

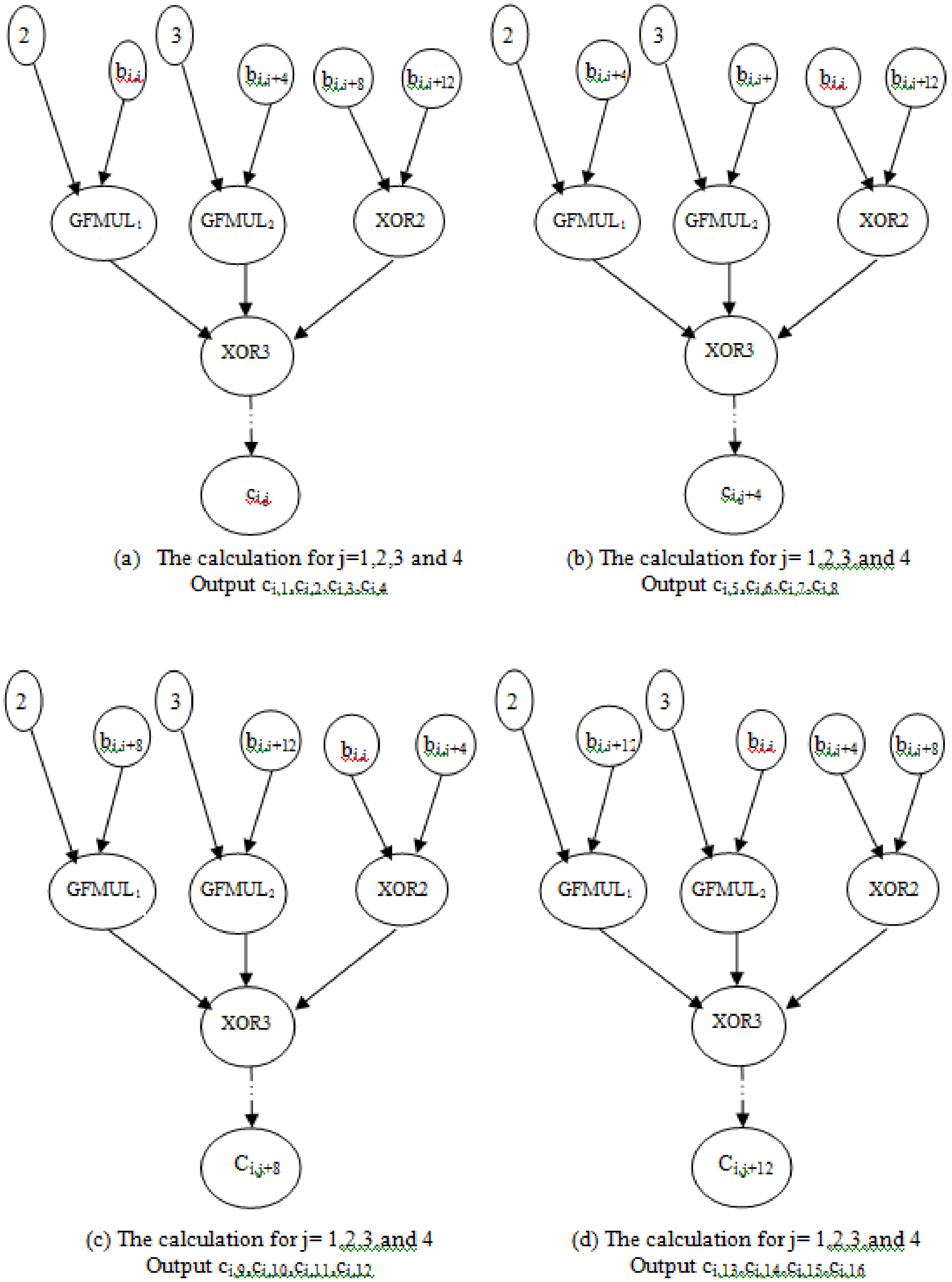

Figure 2: Dataflow graphs for AES algorithm (Encryption Mode): Mix-Column operation

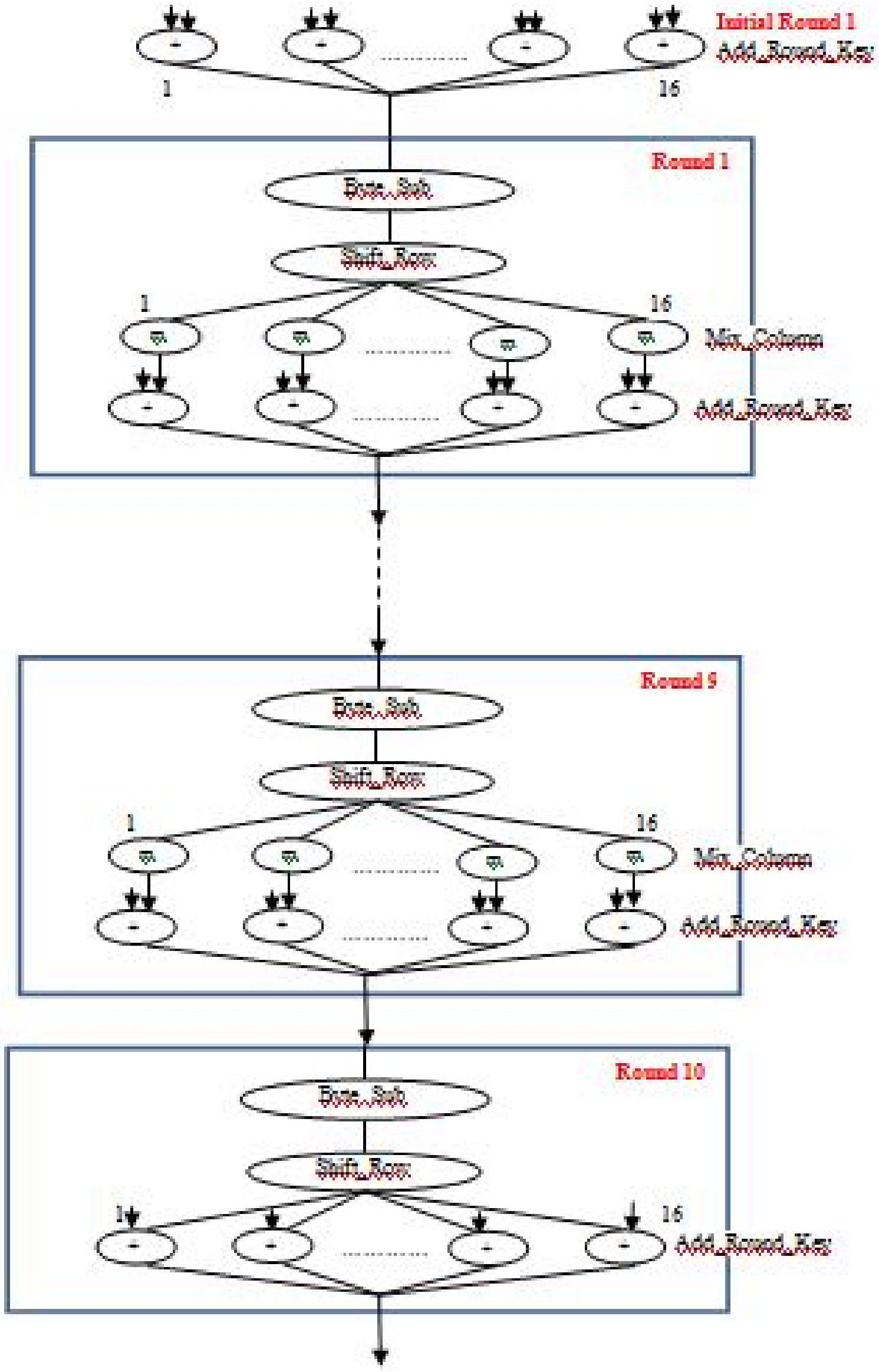

Figure 3: Parallelization of AES encryption operation

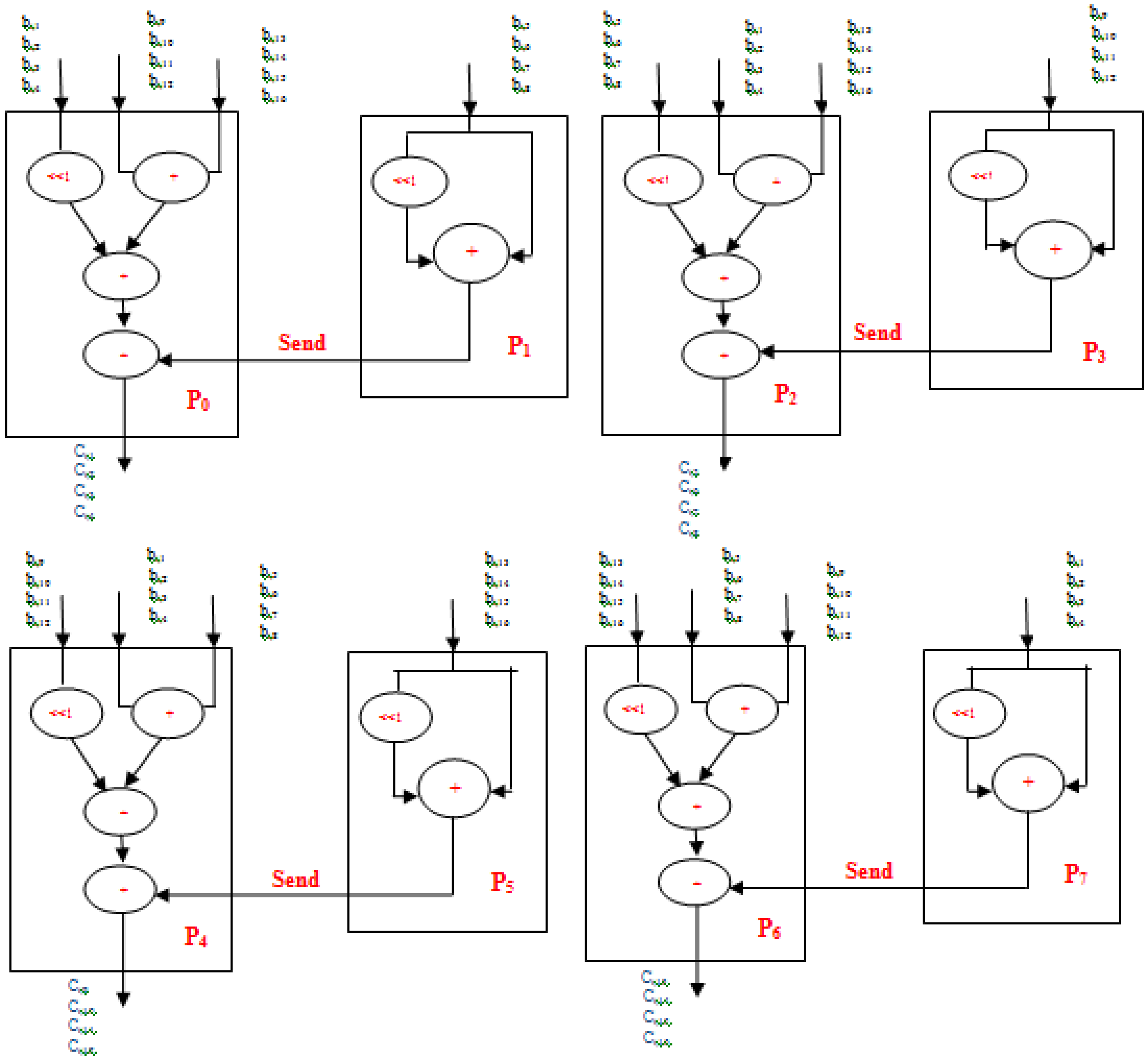

Figure 4: Execute the Mix_Column operation

## 3.2. The Decryption Model

As the encryption operation, the parallelization of decryption can be done in two levels. In the first level of parallelism (pipelining different rounds), the decryption operation is done in the same way as the encryption operation (Section 3.1.1). In order to decrypt *"L"* number of data blocks ( $1 \leq i \leq L$ ), D, $C_i$, and $B_i$, for each block can be represented as follows:

$$C_i = \begin{bmatrix} c_{i,1} & c_{i,2} & c_{i,3} & c_{i,4} \\ c_{i,5} & c_{i,6} & c_{i,7} & c_{i,8} \\ c_{i,9} & c_{i,10} & c_{i,11} & c_{i,12} \\ c_{i,13} & c_{i,14} & c_{i,15} & c_{i,16} \end{bmatrix} \quad \text{and} \quad D = \begin{bmatrix} 0E & 0B & 0D & 09 \\ 09 & 0E & 0B & 0D \\ 0D & 09 & 0E & 0B \\ 0B & 0D & 09 & 0E \end{bmatrix}$$

$$B_i = D * C_i \qquad (7)$$

$$\begin{bmatrix} b_{i,1} & b_{i,2} & b_{i,3} & b_{i,4} \\ b_{i,5} & b_{i,6} & b_{i,7} & b_{i,8} \\ b_{i,9} & b_{i,10} & b_{i,11} & b_{i,12} \\ b_{i,13} & b_{i,14} & b_{i,15} & b_{i,16} \end{bmatrix} = \begin{bmatrix} 0E & 0B & 0D & 09 \\ 09 & 0E & 0B & 0D \\ 0D & 09 & 0E & 0B \\ 0B & 0D & 09 & 0E \end{bmatrix} * \begin{bmatrix} c_{i,1} & c_{i,2} & c_{i,3} & c_{i,4} \\ c_{i,5} & c_{i,6} & c_{i,7} & c_{i,8} \\ c_{i,9} & c_{i,10} & c_{i,11} & c_{i,12} \\ c_{i,13} & c_{i,14} & c_{i,15} & c_{i,16} \end{bmatrix} \quad (8)$$

Inv_Mix_Column is then represented by the following set of equations and illustrated in Figure 5:

$$b_{i,1} = (0E \bullet c_{i,1}) \oplus (0B \bullet c_{i,5}) \oplus (0D \bullet c_{i,9}) \oplus (09 \bullet c_{i,13}) \quad (9)$$

$$b_{i,5} = (09 \bullet c_{i,1}) \oplus (0E \bullet c_{i,5}) \oplus (0B \bullet c_{i,9}) \oplus (0D \bullet c_{i,13}) \quad (10)$$

$$b_{i,9} = (0D \bullet c_{i,1}) \oplus (09 \bullet c_{i,5}) \oplus (0E \bullet c_{i,9}) \oplus (0B \bullet c_{i,13}) \quad (11)$$

$$b_{i,13} = (0B \bullet c_{i,1}) \oplus (0D \bullet c_{i,5}) \oplus (09 \bullet c_{i,9}) \oplus (0E \bullet c_{i,13}) \quad (12)$$

This is repeated for the other three columns of the matrix. As mentioned earlier, both Inv_Mix_Column and Add_Round_key transforms can be executed in parallel. Inv_Mix_Column transformation consists of 160 XOR operations and 192 shift operations. Similar to Mix_Columnu matrix, the elements of Inv_Mix_Column matrix can be computed independently. The Inv_Mix_Column transformation can be executed by at most 64 processors in each stage. Figure 6 describes the details of computing each matrix element in parallel when using 16 processors. As shown in this figure, four processors can cooperate to compute one or more element $b_{i,j}$. In the next section, analysis of the proposed design is detailed.

## 4. ANALYSIS OF THE PROPOSED PARALLEL AES DESIGN

In this section, for both encryption and decryption operations, we discuss the mathematical derivation of the proposed parallel AES design on a pipeline architecture of eleven stages. In our design, "$M_r$" processing elements cooperate to execute each stage (as discussed in Section 3). For simplicity, we assume a block and key sizes of 128 bits.

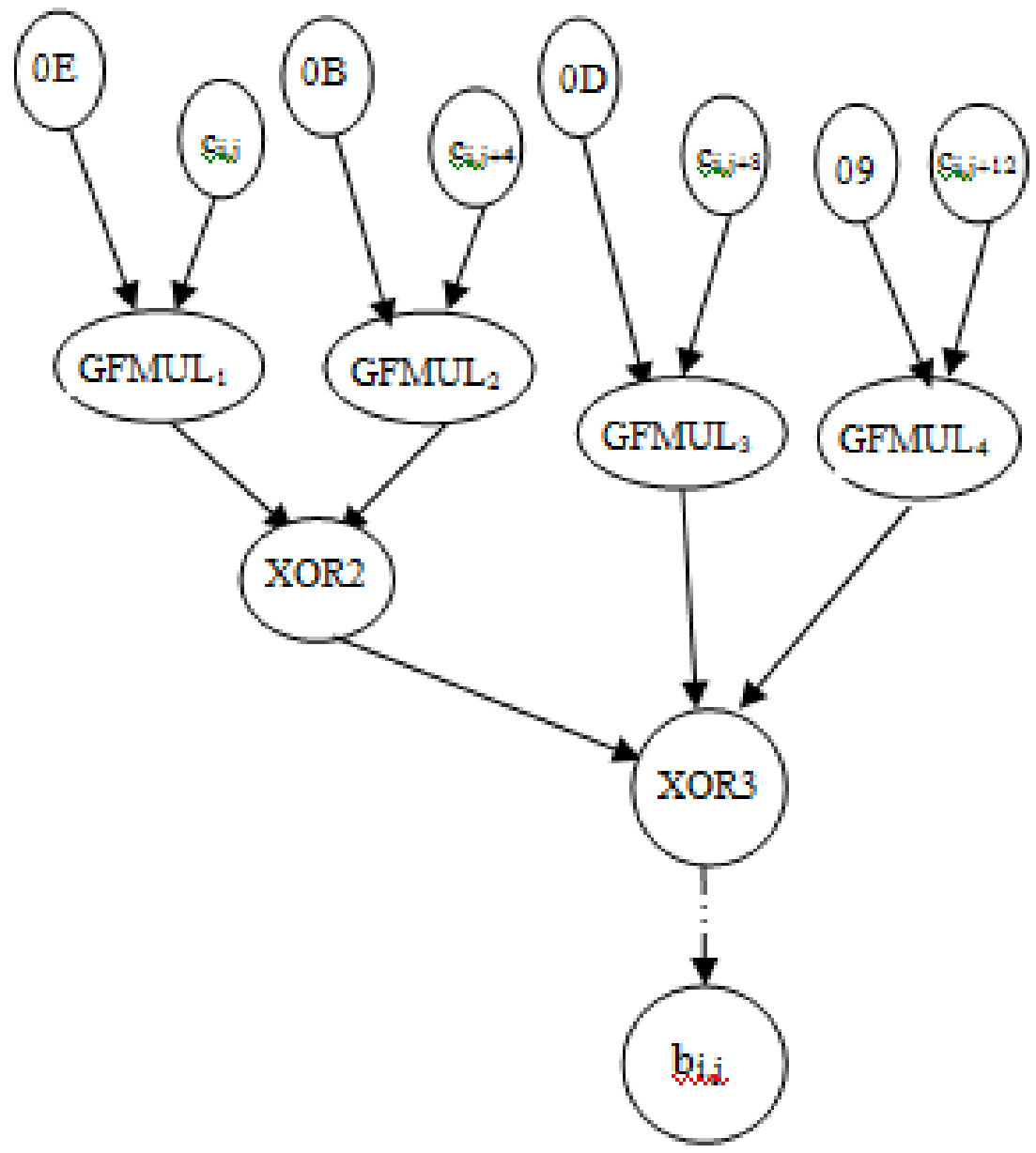

(a) The calculation for j=1,2,3 and 4
Output $b_{i,1}, b_{i,2}, b_{i,3}, b_{i,4}$

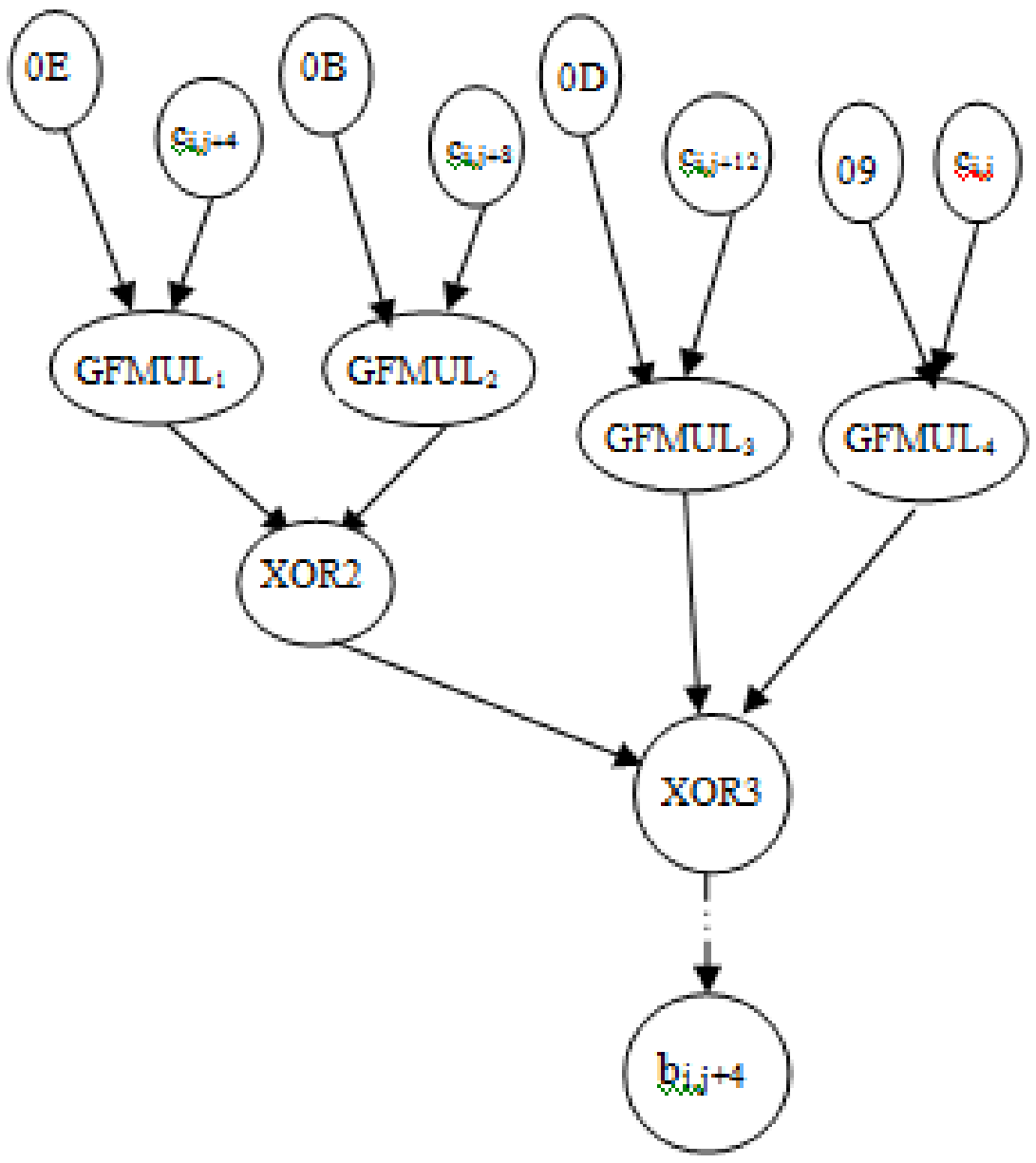

(b) The calculation for j= 1,2,3,and 4
Output $b_{i,5}, b_{i,6}, b_{i,7}, b_{i,8}$

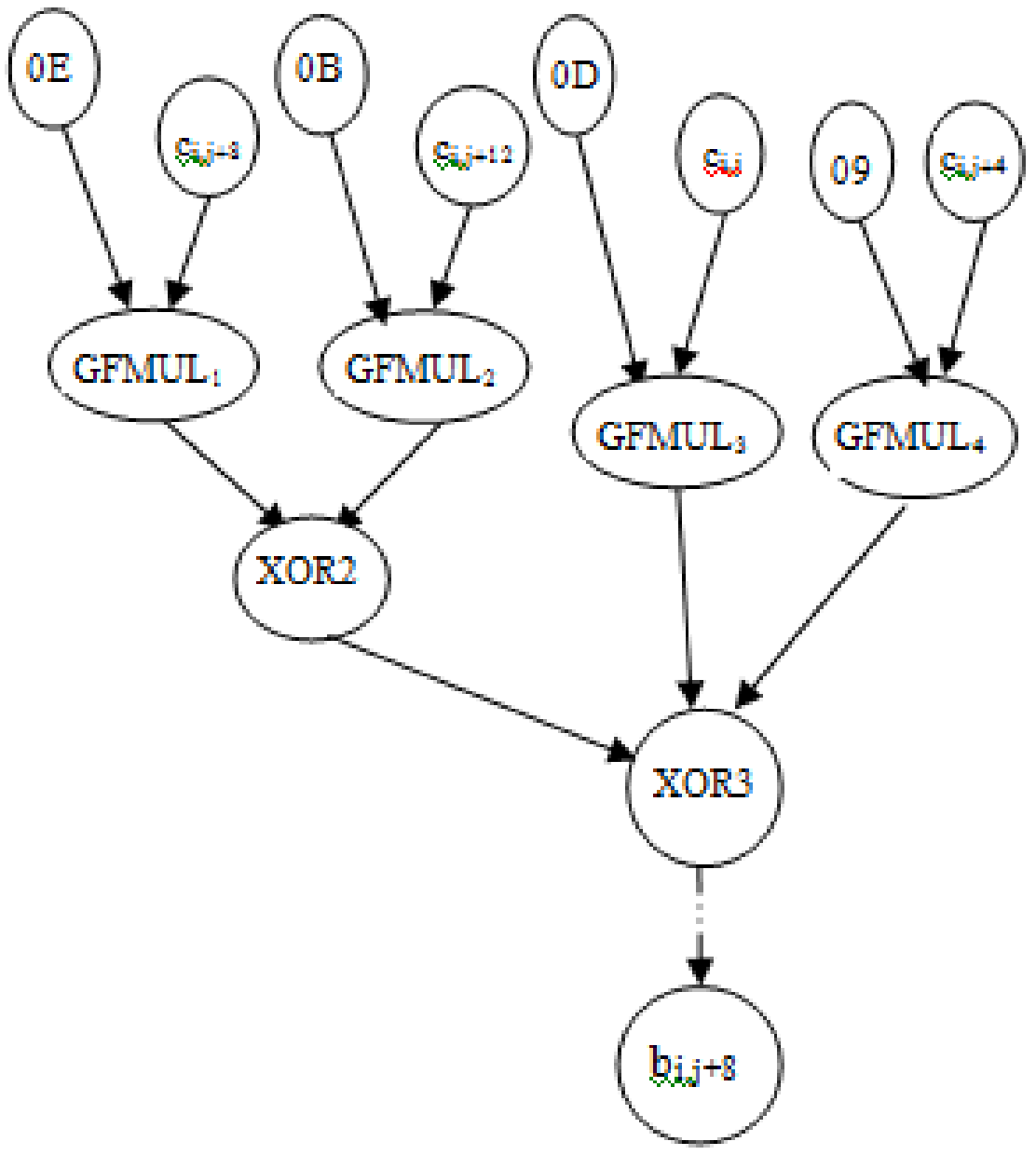

(c) The calculation for j=1,2,3 and 4
Output $b_{i,9}, b_{i,10}, b_{i,11}, b_{i,12}$

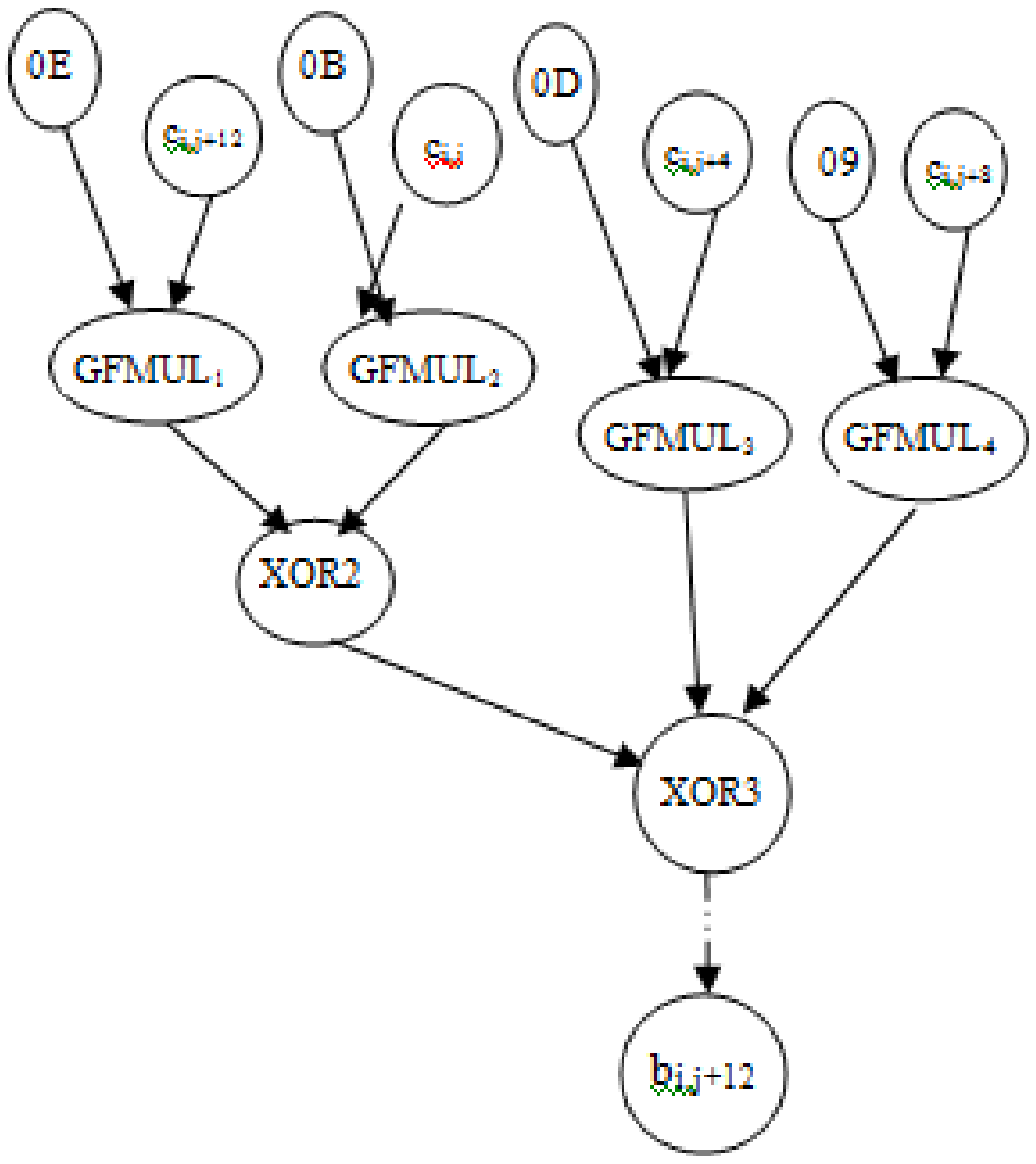

(d) The calculation for j= 1,2,3,and 4
Output $b_{i,13}, b_{i,14}, b_{i,15}, b_{i,16}$

Figure 5: Dataflow graphs for AES algorithm (decryption mode) Inv_Mix_Column operation

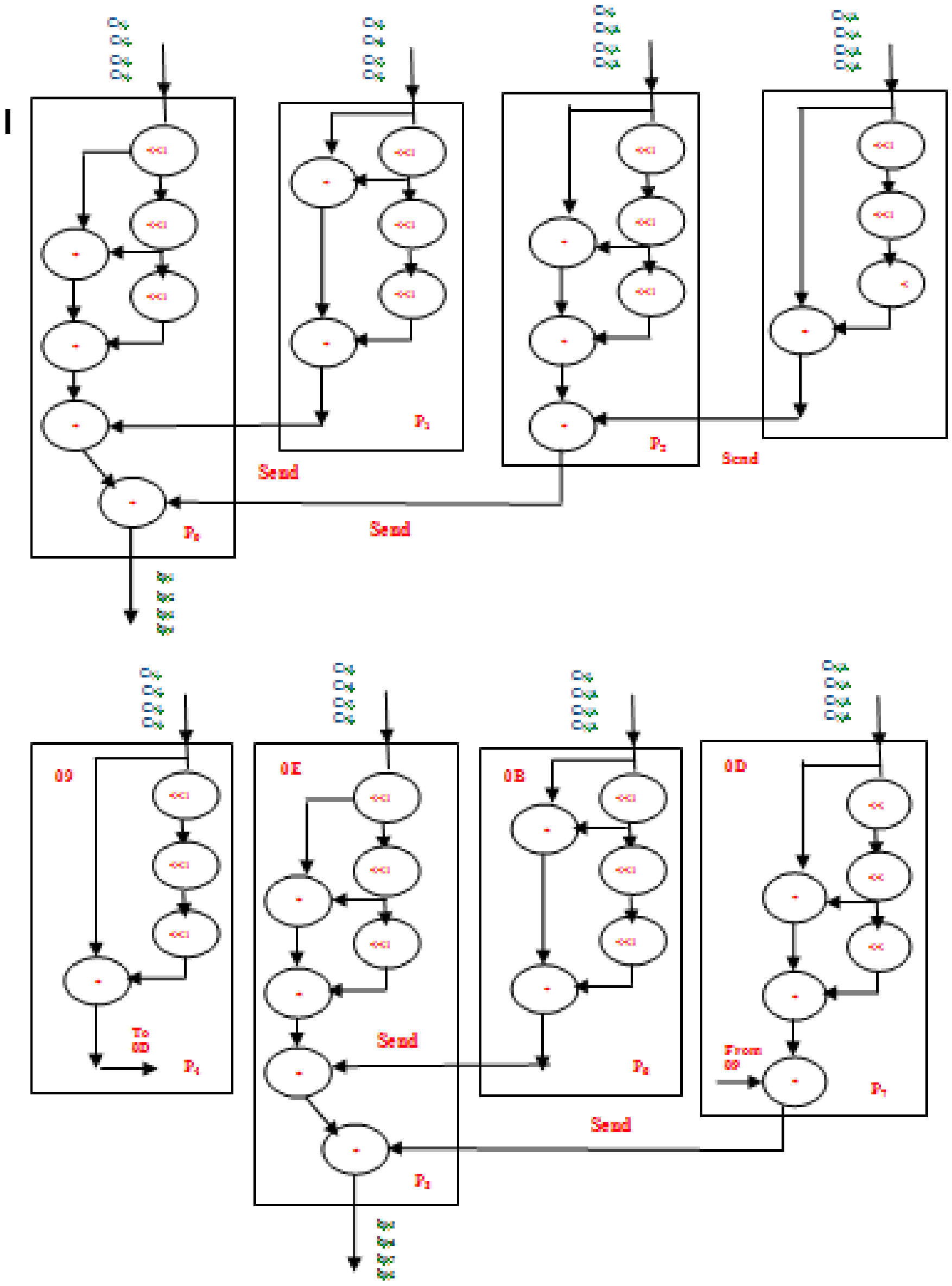

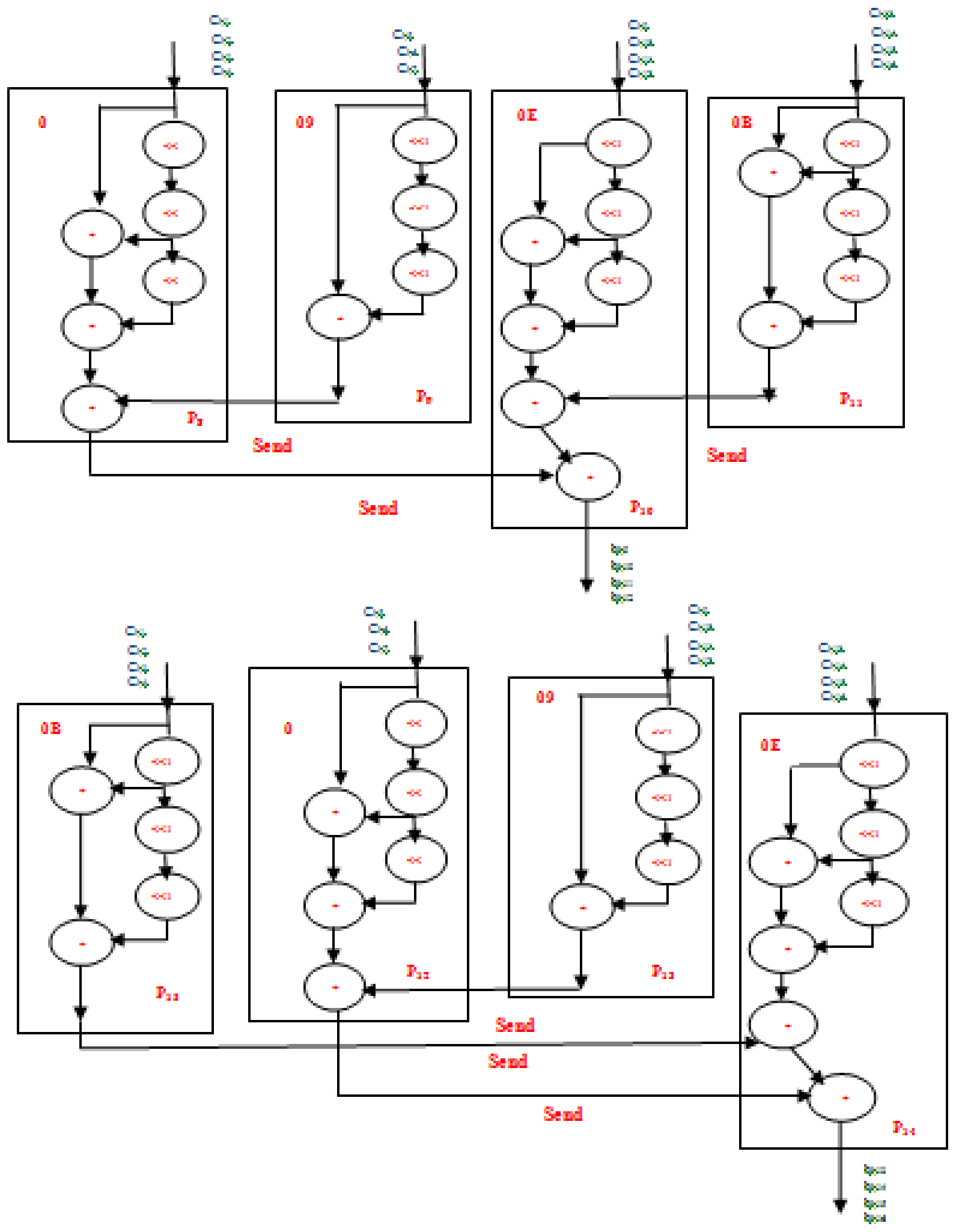

Figure 6: Execute the Inv_Mix_Column operation

## 4.1. Encryption Operation

The total sequential time "$T_{ES}$" needed to execute the encryption operation is given by:

$$T_{ES} = T_{Add_Round_Key} + \{(N_r\text{-}1) * T_{Nr\text{-}1}\} + T_{Nr} \quad (13)$$

Where

$$T_{Nr\text{-}1} = T_{Byte_Sub} + T_{Shift_Row} + T_{Mix_Column} + T_{Add_Round_Key} \quad (14)$$

$$T_{Nr} = T_{Byte_Sub} + T_{Shift_Row} + T_{Add_Round_Key} \quad (15)$$

$$T_{Add_Round_Key} = 16 * T_{XOR} \quad (16)$$

($T_{XOR}$: the time needed to execute one XOR operation)

$$T_{Shift_Row} = 48 * T_{shift} \quad (17)$$

($T_{shift}$ : the time needed to execute one shift operation)

$$T_{Mix_Column} = 16 * (2*T_{shift} + 4*T_{XOR}) \quad (18)$$

$$T_{XOR} = 6 * T_{shift} \quad (19)$$

From equations (13, 14, 15, 16, 17, 18, and 19), we deduce:

$$T_{ES} = 880*T_{XOR} + 10\ T_{Byte_Sub} \quad (20)$$

$T_{Byte_Sub}$ is very small and can be neglected. For "*L"* blocks, the total sequential time is given by:

$$T_{LES} = L* T_{ES} = 880*L*T_{XOR} \quad (21)$$

### 4.1.1. Pipelining the AES encryption rounds

Assuming that the total number of processing elements $M = 11* M_r$, and "$M_r$" processing elements are used to execute each stage, the pipeline time "$T_{pipeline}$" is given by:

$$T_{pipeline} = L*t_1 + 9*t_2 + t_3 \quad (22)$$

$$T_{pipeline} = L*\left( t_1 + \frac{9}{L}*t_2 + \frac{1}{L}*t_3 \right) \quad (23)$$

Where

$t_1$: is the time needed to execute the initial round
$t_2$:is the time needed to execute any round $N_j$ ($1 \le j \le 9$)
$t_3$: is the time needed to execute the final round

$$t_1 = T_{Add_Round_Key} = 16* T_{XOR} \quad (24)$$

$$t_2 = T_{Byte_Sub} + T_{Shift_Row} + T_{Mix_Column} + T_{Add_Round_Key}$$
$$= T_{Byte_Sub} + 48* T_{Shift} + 16 * (2*T_{shift} + 4*T_{XOR}) + 16* T_{XOR} \quad (25)$$

$$t_3 = T_{Byte_Sub} + T_{Shift_Row} + T_{Add_Round_Key}$$
$$= T_{Byte_Sub} + 48* T_{Shift} + 16* T_{XOR} \quad (26)$$

$T_{Byte_Sub}$ is very small and can be neglected. From Eqs. (22 to 26), the pipelined time "$T_{pipeline}$" is given by:

$$T_{pipeline} = L*\begin{pmatrix} 16*T_{XOR} + \frac{10}{L}\left([48*T_{shift}] + [16*T_{XOR}]\right) + \\ \frac{9}{L}*\left(32*T_{shift} + 64*T_{XOR}\right) \end{pmatrix} \quad (27)$$

#### 4.1.2. Parallelization of Add_Round_Key and Mix_Column transformations

We assume that the total number of PEs that compute each round equals to "$M_r$", where $2 \le M_r \le 32$. Therefore, the time needed to execute Add_Round_key transformation is given by:

$$T_{Add_Round_Key} = \left(\frac{16}{M_r}\right) * T_{XOR} \quad (28)$$

While, the time needed for execute Mix_Column transformation is given by:

$$T_{Mix_Column} = \left(\frac{32}{M_r}\right) * \left[Max\left(T_{PEk}, T_{PEk+1}\right) + T_{ov}\right] \quad (29)$$

Where

$$T_{PEk} = (T_{shift} + 3*T_{XOR}) \quad (30)$$

$$T_{PEk+1} = (T_{shift} + T_{XOR})$$
(31)

$T_{ov}$ = the overhead Time

## 4.2. Decryption Operation

The total sequential time "$T_{DS}$" needed to execute the decryption operation is given by:

$$T_{DS} = T_{Add_Round_Key} + (N_r-1) * T_{Nr-1} + T_{Nr} \quad (32)$$

Where

$$T_{Nr-1} = T_{Inv_Byte_Sub} + T_{Inv_Shift_Row} + T_{Inv_Mix_Column} + T_{Add_Round_Key} \quad (33)$$

$$T_{Nr} = T_{Inv_Byte_Sub} + T_{Inv_Shift_Row} + T_{Add_Round_Key} \quad (34)$$

$$T_{Inv_Mix_Column} = 16 * (12*T_{shift} + 10*T_{XOR}) \quad (35)$$

By using the same assumptions at Eqs. (16-19), and from Eqs. (32-35), we deduce:

$$T_{DS} = 1984*T_{XOR} + 10\, T_{Inv_Byte_Sub} \quad (36)$$

$T_{Inv_Byte_Sub}$ is very small and can be neglected. For "$L$" blocks, the total sequential time is given by:

$$T_{LDS} = L*T_{DS} = 1984*L*T_{XOR} \quad (37)$$

### 4.2.1. Pipelining the AES decryption rounds

As discussed in the previous subsection (encryption case), the pipeline time is given by:

$$T_{pipeline} = L*t_4 + 9*t_5 + t_6 \quad (38)$$

$$T_{pipeline} = L*\left(t_4 + \frac{9}{L}*t_5 + \frac{1}{L}*t_6\right) \quad (39)$$

Where

$t_4$: is the time needed to execute the initial round
$t_5$: is the time needed to execute any round $N_j$ $(1 \leq j \leq 9)$
$t_6$: is the time needed to execute the final round

$$t_4 = T_{Add_Round_Key} = 16*T_{XOR} \quad (40)$$

$$t_5 = T_{Inv_Byte_Sub} + T_{Inv_Shift_Row} + T_{Inv_Mix_Column} + T_{Add_Round_Key}$$

$$= T_{Inv_Byte_Sub} + 48 * T_{Shift} + (192 * T_{shift} + 160 * T_{XOR}) + 16 * T_{XOR} \quad (41)$$

$$t_6 = T_{Inv\text{-}Byte_Sub} + T_{Inv_Shift_Row} + T_{Add_Round_Key}$$

$$= T_{Inv_Byte_Sub} + 48 * T_{Shift} + 16 * T_{XOR} \quad (42)$$

From Eqs. (37-42) and with the assumption that $T_{Inv_Byte_Sub}$ is very small and can be neglected, the pipelined time "$T_{pipeline}$" is given by:

$$T_{pipeline} = L * \begin{pmatrix} 16 * T_{XOR} + \frac{10}{L}\left(\left[48 * T_{shift}\right] + \left[16 * T_{XOR}\right]\right) \\ + \frac{9}{L} * \left(192 * T_{shift} + 160 * T_{XOR}\right) \end{pmatrix} \quad (43)$$

### 4.2.2. Parallelization of Add_Round_Key and Inv_Mix_Column transformations

Assuming that the total number of PEs that compute each round equals "$M_r$", where $4 \leq M_r \leq 64$, therefore, the total time for Add_Round_Key transformation is given by:

$$T_{Add_Round_Key} = \left(\frac{16}{M_r}\right) * T_{XOR} \quad (44)$$

While, the time needed to execute the Inv_Mix_Column transformation is given by:

$$T_{Inv_Mix_Column} = \left(\frac{64}{M_r}\right) * \left[Max\left(T_{PEk}, T_{PEk+1}, T_{PEk+2}, T_{PEk+3}\right) + T_{ov}\right] \quad (45)$$

Where $T_{PEk} = (3 * T_{shift} + 4 * T_{XOR})$ (46)

$$T_{PEk+1} = (3 * T_{shift} + 2 * T_{XOR}) \quad (47)$$

$$T_{PEk+2} = (3 * T_{shift} + 3 * T_{XOR}) \quad (48)$$

$$T_{PEk+3} = (3 * T_{shift} + T_{XOR}) \quad (49)$$

$T_{ov}$ = the overhead time

## 4.3. Discussion of Results

In literature, there are some metrics [20] used to evaluate the system performance such as:

- *Execution time (parallel time)* $T_{par}$ is referred to the total running time of the program.
- *Speedup* $S_p$, which relates the time taken to solve the problem on a single processor machine to the time taken to solve the same problem using parallel implementation.
- *Efficiency, Ep,* is defined as the ratio $S_p/M$.
- *Degree of improvement* is the percentage of improvement in system performance with respect to sequential execution and can be determined by $(T_s - T_{par})/T_s$.

Tables 1 and 3 illustrate the improvement of the proposed design with respect to the sequential model for both encryption and decryption operations. On the other hand, Tables 2, and 4 show the effect of parallelization of both Add_Round_Key and Mix_Column/ Inv_Mix_Column transformations for the cases of encryption and decryption respectively for different number of blocks (L = 10, 25, and 40).

Table (1): Degree of improvement with respect to sequential time ($T_{LES}$= L*880*$T_{XOR}$)

(a) Pipelined encryption without parallelization of Add_Round_Key and Mix_Column

| Number of blocks | Sequential time | Pipelining time | Degree of improvement |
|---|---|---|---|
| L=10 | 8800*$T_{XOR}$ | 1024*$T_{XOR}$ | 88% |
| L=25 | 22000*$T_{XOR}$ | 1262.5*$T_{XOR}$ | 94.2% |
| L=40 | 35200*$T_{XOR}$ | 1556*$T_{XOR}$ | 95.5% |

(b) Pipelined encryption with parallelization of Add_Round_Key and Mix_Column

| $M_r$ | L=10 | L=25 | L=40 |
|---|---|---|---|
| 2 | 92% | 96% | 97% |
| 4 | 95% | 97% | 98.5% |
| 8 | 97.3% | 98.8% | 98.9% |

Table (2): The effect of parallelization of Add_Round_Key and Mix_Column

(a) L=10

| Mr | Execution time | Speedup | Efficiency | Degree of improvement |
|---|---|---|---|---|
| 1 | 1022*$T_{XOR}$ | 1 | 1 | - |
| 2 | 696*$T_{XOR}$ | 1.47 | 0.73 | 32% |
| 4 | 388*$T_{XOR}$ | 2.63 | 0.65 | 62% |
| 8 | 234*$T_{XOR}$ | 4.34 | 0.53 | 77% |

(b) L=25

| $M_r$ | Execution time | Speed up | Efficiency | Degree of improvement |
|---|---|---|---|---|
| 1 | 1264*$T_{XOR}$ | 1 | 1 | - |
| 2 | 815*$T_{XOR}$ | 1.55 | 0.77 | 35% |
| 4 | 447.5*$T_{XOR}$ | 2.88 | 0.705 | 65% |
| 8 | 264*$T_{XOR}$ | 4.78 | 0.59 | 79% |

(c) L=40

| $M_r$ | Execution time | Speed up | Efficiency | Degree of improvement |
|---|---|---|---|---|
| 1 | 1556*$T_{XOR}$ | 1 | 1 | - |
| 2 | 936*$T_{XOR}$ | 1.66 | 0.89 | 39.8% |
| 4 | 508*$T_{XOR}$ | 3.06 | 0.76 | 67.2% |
| 8 | 292*$T_{XOR}$ | 5.3 | 0.66 | 81.2% |

Table (3): Degree of improvement with respect to sequential time ($T_{LDS}$= L*1984*$T_{XOR}$)

(a) Pipelined decryption without parallelization of Add_Round_Key and Inv_Mix_Column

| Number of blocks | Sequential time | Pipelining time | Degree of improvement |
|---|---|---|---|
| L=10 | 19840*$T_{XOR}$ | 1984*$T_{XOR}$ | 90% |
| L=25 | 49600*$T_{XOR}$ | 2224*$T_{XOR}$ | 95.5% |
| L=40 | 79360*$T_{XOR}$ | 2464*$T_{XOR}$ | 96.8% |

(b) Pipelined encryption with parallelization of Add_Round_Key and Mix_Column

| $M_r$ | L=10 | L=25 | L=40 |
|---|---|---|---|
| 2 | 93.7% | 97.2% | 98% |
| 4 | 95.9% | 98.2% | 98.8% |
| 8 | 97.7% | 98.9% | 99% |

Table (4): The effect of parallelization of Add_Round_Key and Inv_Mix_Column

(a) L=10

| $M_r$ | Execution time | Speedup | Efficiency | Degree of improvement |
|---|---|---|---|---|
| 1 | 1984*$T_{XOR}$ | 1 | 1 | - |
| 2 | 1248 $T_{XOR}$ | 1.59 | 0.79 | 37.1% |
| 4 | 804 $T_{XOR}$ | 2.46 | 0.61 | 59.5% |
| 8 | 444* $T_{XOR}$ | 4.46 | 0.56 | 77.6% |
| 16 | 262 $T_{XOR}$ | 7.57 | 0.47 | 86.7% |

(b) L=25

| $M_r$ | Execution time | Speedup | Efficiency | Degree of improvement |
|---|---|---|---|---|
| 1 | 2224*$T_{XOR}$ | 1 | 1 | - |
| 2 | 1368 $T_{XOR}$ | 1.62 | 0.81 | 38.5% |
| 4 | 864$T_{XOR}$ | 2.57 | 0.64 | 61.2% |
| 8 | 474 $T_{XOR}$ | 4.71 | 0.59 | 78.7% |
| 16 | 277 $T_{XOR}$ | 8.02 | 0.50 | 87.5% |

(c) L=40

| $M_r$ | Execution time | Speedup | Efficiency | Degree of improvement |
|---|---|---|---|---|
| 1 | 2464*$T_{XOR}$ | 1 | 1 | - |
| 2 | 1488 $T_{XOR}$ | 1.66 | 0.81 | 39.6% |
| 4 | 924 $T_{XOR}$ | 2.67 | 0.67 | 62.5% |
| 8 | 504 $T_{XOR}$ | 4.88 | 0.61 | 79.5% |
| 16 | 292 $T_{XOR}$ | 8.43 | 0.51 | 88.1% |

From the above tables, the following facts could be deduced:

- Tables 1(a) and 3(a) show that using pipeline increases significantly the system performance for the cases of encryption and decryption. In addition, as the number of blocks increases, for cases of encryption and decryption, the degree of improvement increases.
-
- As shown in Table 1(b) and 3(b), as the number of processors used to execute each stage (2 to 16) increases, the improvement degree increases irrespective of the block size. To obtain a reasonable efficiency, we will be satisfied with an improvement degree equals to 98%. Which can be satisfied when $M_r$ =8 for the encryption case and $M_r$ =16 for the decryption case.
- Tables 2 and 4 show the effect of parallelizing Add_Round_Key and Mix_Column/ Inv_Mix_Column transformations on the system performance inside each stage. The comparison with the case of using only one processor is illustrated. As the number of processors increases, the total execution time decreases. In addition, the speedup increases for both encryption and decryption operations. Moreover, the improvement degree increases irrespective of the block size. This is true for L =10, 25, and 40. This leads to the conclusion that the proposed design is scalable and is suitable for real-time applications.

Previous work proposed for pipelining AES algorithm was based on using nine stages. In our work, we propose the use of eleven stages in order to exploit the sources of parallelism in both initial and final round. This enhances the system performance compared to previous designs. In addition, we use two-levels of parallelism: the first level is pipelining different rounds (from round zero to round 10), while the second one is through parallelization both the Add_Round_Key and the Mix-Column transformations. Using two-levels of parallelization benefits from the highly independency of Mix_Column/Inv_Mix_Colum transformation which leads to a better performance.

## 5. CONCLUSIONS

The Advanced Encryption Standard (AES) algorithm is a symmetric block cipher which operates on a sequence of blocks each consists of 128, 192 or 256 bits. Moreover, the cipher key for the AES algorithm is a sequence of 128, 192 or 256 bits. AES algorithm has many sources of parallelism. In this work we proposed an optimized version of AES algorithm. Both the encryption and the decryption algorithms have been optimized. In the present paper, we detailed a design for implementation of AES algorithm on a multiprocessor platform. While most of the previous designs either use pipelined parallelization or take advantage of the Mix_Column parallelization, our design is based on combining pipelining of rounds and parallelization of Mix_Column and Add_Round_Key transformations. This model is divided into two levels: the first one is pipelining different rounds, while the second one is through parallelization of both the Add_Round_Key and the Mix_Column transformations. Previous work proposed for pipelining AES algorithm was based on using nine stages, while, we propose the use of eleven stages in order to exploit the sources of parallelism in both initial and final round. This enhances the system performance compared to previous designs. Using two-levels of parallelization benefits from the highly independency of Add_Round_Key and Mix_Column/ Inv_Mix_Colum transformations. The analysis shows that using pipeline increases significantly the degree of improvement for both encryption and decryption by approximately 95%. Moreover, parallelizing Add_Round_Key and Mix_Column/ Inv_Mix_Column transformations increases the degree of improvement by approximately 98%. To obtain a reasonable efficiency, we will be satisfied with an improvement degree equals to 98%. This could be achieved using eight processors for each stage in case of encryption and sixteen processors for the decryption case. Since, the increase of number of processors will decrease the efficiency. The analysis shows that the improvement degree increases irrespective of the block size. This is true for L =10, 25, and 40. This leads to the conclusion that the proposed design is scalable and is suitable for real-time applications.

## REFRENCES

[1] Joan Daemen and Vincent Rijmen, (1998) "AES Proposal: Rijndael"

[2] W.Stallings (2010), Cryptography and Network Security: Principles and Practice, Prentice Hall.

[3] Mostafa I. Soliman and Ghada Y. Abozaid, (2010) "FastCrypto: Parallel AES Pipelines Extension for General-Purpose Processors", Neural, Parallel, and Scientific Computations, No. 18, pp. 47 – 58.

[4] S.-M. Yoo, D. Kotturi, D.W. Pan, and J. Blizzard, (2005) "An AES crypto chip using a high-speed parallel pipelined architecture", Microprocessors and Microsystems, No.29, pp. 317–326.

[5] A.Hodjat, and I. Verbauwhede (2004), "A 21.54 Gbits/s fully pipelined AES processor on FPGA", in Proc. of 12th Annual IEEE Symposium on Field-Programmable Custom Computing Machines (FCCM'04), pp. 308-309
[6] Bin Liu, and Bevan M. Baas (2013) "Parallel AES Encryption Engines for Many-Core Processor Arrays", IEEE Transactions on Computers, Vol. 62, no. 3, pp. 536-547.
[7] C.Ananth and K. Ramu (2008) "Fully pipelined implementations of AES with speeds exceeding 20 Gbits/s with S-boxes implemented using logic only", Technical report Department of ECE, George Mason University.
[8] Y.Mitsuyama, M. Kimura, T. Onoye, and I. Shirakawa, (2005) "Architecture of IEEE802.11i Cipher Algorithms for Embedded Systems", IEICE Transactions on Fundamentals of Electronics, Communications and Computer Sciences, Vol. E88-A, no.4, pp.899-906.
[9] S.Arrag, A. Hamdoun, A. Tragha and S. Khamlich, (2012) "Design and Implementation A different Architectures of Mix_Column in FPGA", International Journal of VLSI Design and Communication Systems, Vol. 3, Issue 4, p.11.
[10 ]M.Anitha and S. Priya, (2014) "Design of Low Power Mixcolumn in Advanced Encryption Standard Algorithm", International Journal of Scientific and Engineering Research (IJSER), Vol. 5, Issue 4, pp. 64-68.
[11] P.Noo-intara, S. Chantarawong, and S. Choomchuay, (2004) "Architectures for MixColumn Transform for the AES", in Proc. of ICEP 2004, Phuket, Thailand, pp. 152-156.
[12] Sabbir Mahmud, (2004) "A Study on Parallel Implementation of Advanced Encryption Standard (AES)", M.S. thesis, Computer Science, Independent University, Bangladesh, May, 2004.
[13] C.So-In, S. Poolsanguan, C. Poonriboon, K. Rujirakul, and C. Phudphut, (2013) "Performance Evaluation of Parallel AES Implementations over CUDAGPU Framework", International Journal of Digital Content Technology and its Applications (JDCTA), Vol.7, no.5, pp. 501-511.
[14] S.Tillich, and J. Großschädl, (2006) "Instruction Set Extensions for Efficient AES Implementation on 32-bit Processors, Cryptographic Hardware and Embedded Systems (CHES)", Lecture Notes in Computer Science, Vol.4249, pp 270-284.
[15] A.Elbirt, (2007) "Fast and Efficient Implementation of AES Via Instruction Set Extensions", in Proc. of the 21st International Conference on Advanced Information Networking and Applications Workshops (AINAW'07), Niagara Falls, Ont. , 21-23 May, Vol.1, pp. 396-403.
[16] S.Gueron, (2012) "Intel® Advanced Encryption Standard (AES) Instructions Set. Intel", White Paper, "https://software.intel.com/en-us/articles/intel-advanced-encryption-standard-aes-instructions-set"
[17] Brisk, A. Kaplan, and M. Sarrafzadeh (2003), "Parallel Analysis of the Rijndael Block Cipher", in Proc. of the IASTED International Conference of Parallel and Distributed Computing and Systems, Marina del Rey, USA, 3-5 Nov.
[18] Jung Ho Yoo, (2011) "Fast Software Implementation of AES-CCM on Multiprocessors", Algorithms and Architectures for Parallel Processing, Lecture Notes in Computer Science, Vol. 7017, pp. 300-311.
[19] M.S. Arun, and V. Saminathan, (2014) "Parallel AES Encryption with Modified Mix-columns For Many Core Processor Arrays", International Journal of Engineering Science and Innovative Technology (IJESIT), Vol. 3, Issue 3, pp. 184-190.
[20] J.Hennessy and D. Patterson, (2003), Computer Architecture: a Quantitative Approach, Morgan Kaufmann Publishers.

## AUTHORS

*Ghada F. ElKabbany* is an Assistant Professor at Electronics Research Institute, Cairo-Egypt. She received her B. Sc. degree, M. Sc. degree and Ph. D. degree in Electronics and Communications Engineering from Faculty of Engineering, Cairo University, Egypt in 1990, 1994 and 2007 respectively. Her research interests include: High Performance Computing (HPC), computer network security, robotics, and image processing.

*Heba K. Aslan* is a Professor at Electronics Research Institute, Cairo-Egypt. She received her B.Sc. degree, M.Sc. degree and Ph.D. degree in Electronics and Communications Engineering from the Faculty of Engineering, Cairo University, Egypt in 1990, 1994 and 1998 respectively. Aslan has supervised several masters and Ph.D. students in the field of computer networks security. Her research interests include: Key Distribution Protocols, Authentication Protocols, Logical Analysis of Protocols and Intrusion Detection Systems.

*Mohamed N. Rasslan* is an Assistant Professor at Electronics Research Institute, Cairo, Egypt. He received the B.Sc., M.Sc., degrees from Cairo University and Ain Shams University, Cairo, Egypt, in 1999 and 2006 respectively, and his Ph.D. from Concordia University, Canada 2010. His research interests include: Cryptology, Digital Forensics, and Networks Security.